\documentstyle [12pt,a4p,epsfig,amsmath,multicol]{article}
\begin{document}
\vspace*{0.6cm}

\begin{center} 
{\normalsize\bf Consequences for special relativity theory of restoring
 Einstein's neglected additive constants in the Lorentz Transformation}
\end{center}
\vspace*{0.6cm}
\centerline{\footnotesize J.H.Field}
\baselineskip=13pt
\centerline{\footnotesize\it D\'{e}partement de Physique Nucl\'{e}aire et 
 Corpusculaire, Universit\'{e} de Gen\`{e}ve}
\baselineskip=12pt
\centerline{\footnotesize\it 24, quai Ernest-Ansermet CH-1211Gen\`{e}ve 4. }
\centerline{\footnotesize E-mail: john.field@cern.ch}
\baselineskip=13pt
\vspace*{0.9cm}
\abstract{ Evaluation of the additive constants in the space-time Lorentz
  transformation equations required, according to Einstein, to correctly 
  describe synchronised clocks at different spatial locations, reveals
  the spurious and unphysical nature of the `relativity of simultaneity' 
   and `length contraction' effects of conventional special relativity.
   Unlike time dilation, there is no experimental evidence for these
   putative effects. Only a universal (position independent) time dilation
    effect for different inertial frames distinguishes special from 
    Galilean relativity.}
\vspace*{0.9cm}
\normalsize\baselineskip=15pt
\setcounter{footnote}{0}
\renewcommand{\thefootnote}{\alph{footnote}}

   In Einstein's seminal 1905 paper on special relativity~\cite{Ein1} can be found,
   just after the derivation of the space-time Lorentz transformation equations, the following
   statement:
\par
\par `Macht man \"{u}ber die Anfanslage des bewegten Systems und \"{u}ber den
  Nullpunkt von  $\tau$ keinerlei Voraussetzung, so ist auf
  den rechten Seiten dieser Gleichungen je eine additive Konstante
  zuzuf\"{u}gen'
\par
\par
    or, in English:
\par
\par  `If no assumption whatever be made as to the initial position of
  the moving system and as to the zero point of $\tau$
   an additive constant is to be placed on the right side of
  these equations'
\par
\par This assertion is shown here to be of crucial importance for a correct understanding
  of the predictions of special relativity. The quantity $\tau$ above is the recorded time (epoch)
   of a clock in uniform motion relative to Einstein's `stationary frame'. 
  \par Considering the Lorentz transformation between events in two inertial frames
   S, ${\rm S}'$ (where ${\rm S}'$ moves with speed $v$ along a common $x$,$x'$ axis while the  $y$,$y'$
  and $z$,$z'$
   axes are also chosen to be parallel) Einstein's assertion, quoted above, requires that, in
   general, the Lorentz transformation equations relating an event with coordinates
   ($x$,$y$,$z$,$t$)
  in S to the corresponding ones ($x'$,$y'$,$z'$,$t'$) in ${\rm S}'$ must be written as:
   \begin{eqnarray}
      x' & = & \gamma [x - v t]+X,  \\ 
      t' & = & \gamma [t - \frac{v x}{c^2}]+T, \\
      y' & = & y +Y, \\
    z' & = & z +Z.
   \end{eqnarray}
    where $\gamma \equiv 1/\sqrt{1-(v/c)^2}$, and $c$ is the speed of light in free space.
     The constants $Y$ and $Z$ depend on the choice of the $y$,$y'$ and $z$,$z'$ coordinate
    systems respectively, while $X$ and $T$ depend both on the choice of $x$,$x'$ coordinate
    systems and the synchronisation constants of the clocks at rest in S and ${\rm S}'$ that register
    the epochs $t$ and $t'$ respectively. What is the operational physical meaning of the space-
    and time-coordinate symbols in Eqs.~(1)-(4)?  To again quote Einstein~\cite{Ein2}:
 \par
   \par `{\it A priori} it is quite clear that we must be able to learn something about 
      the physical behaviour of measuring rods and clocks from the equations of transformation, for
     the magnitudes $x$, $y$, $z$, $t$ are nothing more nor less than results of measurements
     obtainable by means of measuring rods and clocks.' 
 \par
    \par Consider then, a clock, ${\rm C}'$, at rest on the $x'$ axis at a fixed position, $x'({\rm C}') = \chi$,
    that registers the epoch $t'$. The epoch $t$ in (1) and (2) is that registered by a second clock, C,
     at rest at an arbitary position in the frame S. Suppose that, at some given instant in both S and ${\rm S}'$,
     C and ${\rm C}'$ are synchronised so that  $t = \tau$ and $t' = \tau'$. This can conveniently be done by performing
     the synchronisation at the instant when C and ${\rm C}'$ are aligned in $x$~\cite{Ein1}.
     Any values of the synchronisation constants $\tau'$ and $\tau$ ---the clock epochs that are set
     at the instant of synchronisation--- can be chosen without affecting any physical 
      predictions. Also, without any loss of generality for physical
     predictions, the, in general, arbitary spatial coordinate system in S can be chosen so that, at the epoch  $t = \tau$,
     $x({\rm C}') = \chi$. This corresponds to a particular choice of the position of the orgin of
      the $x$-coordinate system in S relative to the position of ${\rm C}'$ at $t = \tau$.  With this set of initial
     conditions, the constants $X$ and $T$ are given by the equations:
    \begin{eqnarray}
     X & = & \chi-\gamma(\chi-c\tau), \\
     T & = & \tau'-\gamma(\tau - \frac{v \chi}{c^2}).
  \end{eqnarray}
     Subsituting these values of the constants into (1) and (2) and rearranging gives:
\begin{eqnarray}
      x'({\rm C}') - \chi & = & \gamma [x({\rm C}')-\chi - v (t-\tau)] = 0,  \\ 
      t'-\tau' & = & \gamma [t -\tau - \frac{v(x({\rm C}')-\chi)}{c^2}].
   \end{eqnarray}
  It can be seen that the physical content of the space transformation equation (7)
   is just a statement of the equations of motion (world lines) of the clock ${\rm C}'$ 
   in the frames ${\rm S}'$ and S as specifed  by the initial conditions:
   \begin{equation}
    x'({\rm C}') = \chi,~~~~ x({\rm C}') = \chi+v(t-\tau) 
   \end{equation}
    which are the same as in Galilean relativity, given by the $c \rightarrow \infty$, $\gamma \rightarrow 1$  limit
    of (7) and (8). 
    \par Using the second member of (7) to eliminate $ x({\rm C}')-\chi$ from (8), and the definition 
      of $\gamma$, gives the time dilation
     relation that connects corresponding time intervals in the frames S and ${\rm S}'$:
   \begin{equation}
     t - \tau = \gamma(t'-\tau').
  \end{equation}
    Notice the important point that this equation contains no spatial coordinates; in
   particular it is independent of the position of ${\rm C}'$ in the frame ${\rm S}'$ specified by the
    parameter $\chi$. The equations
   (9) and (10) have exactly the same physical content as the Lorentz transformation equations
    (7) and (8). Since the space transformation equations in (9) are the same as in Galilean
    relativity, the only change in passing from Galilean to special relativity is to
    replace Newtonian absolute time $t = t'$ by different corresponding time intervals 
     in the frames S and ${\rm S}'$ that are related by the time dilation relation (10).
    Since no spatial coordinates appear in (10), two clocks ${\rm C}'_1$, ${\rm C}'_2$ at arbitary
    positions  in ${\rm S}'$, that are synchronised with C at the common epoch:
     $t = \tau$ so that $t'_1= t'_2 = \tau' $, remain so at all times:
 \begin{equation}
     t - \tau = \gamma(t'_1-\tau')=  \gamma(t'_2-\tau')
  \end{equation}
        so that
  \begin{equation}
       t'_1 = t'_2~~~~({\rm for~all~values~of}~t).    
    \end{equation}  
   There is therefore no `relativity of simultaneity' effect for such clocks when 
   they are observed from the frame S. It is shown in Eq.~(11) that identical clocks,
   at arbitary positions in ${\rm S}'$, all run slow by the same factor, $1/\gamma$, when observed from the frame S. 
   \par A possible  synchronisation procedure is the following:
   The distance
  between ${\rm C}'_1$ and ${\rm C}'_2$ is adjusted so that  ${\rm C}'_1$ (${\rm C}'_2$) arrive 
  simultaneously at $x = \chi_1$ ($x = \chi_2$). This can obviously always
  be done, regardless of whether a putative `length contraction' effect 
  exists or not. The clocks  ${\rm C}'_1$ and ${\rm C}'_2$ are intially stopped and set
  to epoch $\tau'$. A third clock, at rest in S, is initially stopped, set to
  epoch $\tau$, and placed at $x = \chi_1$. The clocks  ${\rm C}'_1$ and ${\rm C}'_2$
  are then set in uniform motion, and at the instant when  ${\rm C}'_1$ is at
  $x = \chi_1$ and   ${\rm C}'_2$ is at $x = \chi_2$, mechanical switches or 
  electronic signals start all three clocks. At any subsequent instant
  Eqs.~(11) and (12) are satisfied.
   \par If the clocks ${\rm C}'_1$ and ${\rm C}'_2$ are positioned at $x'_1 = \chi_1$ and  $x'_2 = \chi_2$
    respectively, the world lines of the clocks in ${\rm S}'$ and S, with the same choice of coordinate
    systems in S and ${\rm S}'$ as in Eq.~(9) are
   \begin{eqnarray}
  x'_1 &  = & \chi_1,~~~~ x_1 = \chi_1+v(t-\tau),  \\
 x'_2 &  = & \chi_2,~~~~ x_2 = \chi_2+v(t-\tau).
  \end{eqnarray}
   It follows from (13) and (14) that
   \begin{equation}
  x'_2- x'_1 \equiv d' = x_2- x_1 \equiv d = \chi_2-\chi_1~~~({\rm for~all~values~of}~t).   
    \end{equation}
    There is therefore no relativistic `length contraction' effect.

     \par The universal time dilation effect for synchronised clocks at different spatial positions
       was exemplified in a CERN experiment in which which muons were created almost 
       simultaneously by pion decay at different positions around a storage ring~\cite{MUTD}.
       The ${\rm S}'$ frame times, $t'$, in this experiment correspond to the previously measured
       rest frame muon mean lifetime, while the S frame time, $t$, was measured by a precise clock
       (`digitron') in the experimental apparatus. In this experiment the time dilation effect of Eq.~(11)
      was verified with a relative precision of 0.1\% for $\gamma = 29.3$.
     \par  The spurious `relativity of simultaneity' and  `length contraction' effects of conventional
      special relativity arise when, following Einstein in the original special relativity paper~\cite{Ein1},
       Lorentz transformations are employed
    without due regard for the correct choice of the constants $X$ and $T$ in Eqs.~(1) and (2) to 
    correctly describe synchronised clocks at at rest at different positions in the frame ${\rm S}'$.
     The Lorentz transformation, as written down by Lorentz and derived by Einstein, has $\chi = X = 0$ and
     $T = \tau = \tau' =  0$ so that the clock ${\rm C}'$, placed at the origin of ${\rm S}'$, is synchronised so that
      $t = t'= 0$ when $x({\rm C}') = 0$. The mistake made by Einstein (in spite of the
     passage from Ref.~\cite{Ein1} quoted above) and to the present writer's best knowledge,
     by all subsequent authors, before the work presented in Ref.~\cite{JHFLLT}, was to assume
  that the same equation with $X=0$, $T = 0$ is also valid to describe a synchronised clock 
   with $x' \ne 0$, for example, with $x' = \chi$. Thus if a synchronised clock at $x'$ is denoted by
  ${\rm C}'(x')$ it is assumed that the same equations that correctly describe such a synchronised clock
    (with $\tau = \tau' = 0$) at $x' = 0$ (i.e. Eqs. (7) and (8) with $\chi = 0$, $\tau = \tau' = 0$):
     \begin{eqnarray}
      x'({\rm C}'(0)) & = & \gamma [x({\rm C}'(0))- v t] = 0, \\ 
      t'({\rm C}'(0)) & = & \gamma [t - \frac{v x({\rm C}'(0))}{c^2}],
    \end{eqnarray}
     also describe such a clock with $x' = \chi$:
     \begin{eqnarray}
      x'({\rm C}'(\chi)) & = & \gamma [x({\rm C}'(\chi ))- v t],     \\ 
      t'({\rm C}'(\chi)) & = & \gamma [t - \frac{v x({\rm C}'(\chi))}{c^2}]. 
    \end{eqnarray}
     Subtracting (16) from (18) gives:
  \begin{equation}   
   \chi = x'({\rm C}'(\chi))-  x'({\rm C}'(0)) \equiv d' = \gamma[x({\rm C}'(\chi ))-x({\rm C}'(0))] \equiv 
       \gamma d 
  \end{equation}
  which is the `length contraction' effect . Subtracting (17) from (19) and using 
  (20) gives
 \begin{equation}  
     t'({\rm C}'(\chi)) -  t'({\rm C}'(0)) = -\frac{v \chi}{c^2} 
 \end{equation}

     which is a `relativity of simultaneity' effect since the unique epoch $t$ in S corresponds to different epochs
       $t'({\rm C}'(0))$ and $t'({\rm C}'(\chi))$ registered by the clocks ${\rm C}'(0)$ and ${\rm C}'(\chi)$ in ${\rm S}'$. 
   These two correlated `effects' are both spurious since the clock ${\rm C}'(\chi)$,
   described by (18) and (19) is not synchronised with  ${\rm C}'(0)$ described by (16) and (17).
   The correct transformation equations to describe a synchronised clock at $x' = \chi$
    with $\tau = \tau' = 0$ are, from (7) and (8):
    \begin{eqnarray}
      x'({\rm C}'(\chi))- \chi& = & \gamma [x({\rm C}'(\chi ))-\chi- v t]  =0,    \\ 
      t'({\rm C}'(\chi)) & = & \gamma \{t - \frac{v[x({\rm C}'(\chi))-\chi]}{c^2}\}
    \end{eqnarray}
    (16) and (22) give, for all values of $t$, 
   \begin{equation} 
    \chi = x'({\rm C}'(\chi))-  x'({\rm C}'(0)) \equiv d' = x({\rm C}'(\chi ))-x({\rm C}'(0)) \equiv d 
    \end{equation}
  consistent with the general relation (15), while combining (16) with (17) and (22) with(23), as in the derivation
    of (10) from (7) and (8), gives the time dilation relations: 
     \begin{equation}
     t = \gamma  t'({\rm C}'(0)) = \gamma  t'({\rm C}'(\chi)) 
     \end{equation}
    consistent with (11) above, and showing no `relativity of simultaneity' effect, in contradiction to (21).
     Comparing (18) and (19) with the general transformation (7) and (8) for a clock at $x' = \chi$, with
    synchronisation constants $\tau$, $\tau'$, it can be seen
    that the synchronisation constants for the clocks C and 
    ${\rm C}'(\chi)$ take, in (18) and (19), the $\chi$-dependent values
     \begin{equation}      
       \tau(\chi) = -\tau'(\chi) = \frac{\chi}{v}\left(1-\frac{1}{\gamma}\right)
   \end{equation}
      instead of $\tau = \tau' = 0$ in (22) and (23). The clock  ${\rm C}'(\chi)$,
     as described by (18) and (19), is therefore synchronised neither with C nor with
       ${\rm C}'(0)$. Clock synchronisation is, however, a purely mechanical or electronic procedure,
     fully under the control of the experimenter, with no relevance to the physics
     of space-time.
    \par The above arguments show that all text books treating special relativity should
    be corrected and hundreds of papers on special relativity in the pedagogical literature
    should be reevalued. 
    \par At the time of this writing, although there is ample experimental verification 
      of time dilation, or, equivalently, the relativistic transverse Doppler effect,
        none at all exists for `relativity of simultaneity' or `length contraction'~\cite{JHFLLT}.
     Experiments using satellites in low Earth orbit, or GPS satellites, sensitive to
     the existence of an O($\beta$) `relativity of simultaneity' effect,
      have recently been proposed by the present author~\cite{JHFSEXPS}.

\end{document}